# Towards Holistic Language-video Representation: the language model-enhanced MSR-Video to Text Dataset

YUCHEN YANG and YINGXUAN DUAN, Swiss Federal Institute of Technology Lausanne, Switzerland

A more robust and holistic language-video representation is the key to pushing video understanding forward. Despite the improvement in training strategies, the quality of the language-video dataset is less attention to. The current plain and simple text descriptions and the visual-only focus for the language-video tasks result in a limited capacity in real-world natural language video retrieval tasks where queries are much more complex. This paper introduces a method to automatically enhance video-language datasets, making them more modality and context-aware for more sophisticated representation learning needs, hence helping all downstream tasks. Our multifaceted video captioning method captures entities, actions, speech transcripts, aesthetics, and emotional cues, providing detailed and correlating information from the text side to the video side for training. We also develop an agent-like strategy using language models to generate high-quality, factual textual descriptions, reducing human intervention and enabling scalability. The method's effectiveness in improving language-video representation is evaluated through text-video retrieval using the MSR-VTT dataset and several multi-modal retrieval models.

CCS Concepts: • **Information systems** → **Information retrieval**; • **Computing methodologies** → **Computer vision**; • **Applied computing** → **Digital libraries and archives**.

Additional Key Words and Phrases: Language-video embedding, video captioning, language model

## 1 INTRODUCTION

Video understanding is becoming fundamental and the next frontier for many real-world applications, including autonomous driving, video retrieval, and the recent advancement in multi-modal generative models. A more robust and holistic language-video representation is the key to pushing video understanding forward.

With multi-modal contrastive learning using heterogeneous web data becoming a success in image-text representation and contrastive learning is now widely used for language-video representation. Such a task normally consists of 4 steps: Video representation extraction; Textual representation extraction; Feature embedding and matching; and feature representation training.

As large-scale pertaining is proven effective for many complex tasks, a rising amount of research focuses on transferring the existing text-image knowledge to improve language-video representation. Works such as CLIP4Clip[22] and CLIP2Video[9], using CLIP[26] for large-scale pretraining, achieve state-of-the-art results on various language-video datasets. Despite the success of CLIP-based models, these methods still treat the language-video representation as

Authors' Contact Information: Yuchen Yang, yuchen.yang@epfl.ch; Yingxuan Duan, yingxuan.duan@epfl.ch, Swiss Federal Institute of Technology Lausanne, Lausanne, Switzerland.

a text-visual task, where the focus is to align the video's visuals to textual information. However, language-video representation is more complex than just that and the multifaceted modality and information from the video side should also be considered.

Another thread of work, such as MMT[10] and HiT[21], picks up specifically the multi-modal information, such as audio, speech, motion for the video representation extraction process. By including multi-modal cues for the embedding on the video's side, these works open up the door for more diverse and detailed information for the overall language-video representation learning.

Moreover, the lack of high-quality language-video datasets is gradually being noticed. Early language-video datasets, such as MSR-VTT[41], use manual annotation. The labour-intensive and expensive annotation consists of only plain and basic descriptions of the most obvious visual elements within the video. Some following datasets, like HowTo100M[24], start to use automatic speech recognition (ASR) to make transcripts and construct the video-text pairs for training. The automated method has greatly enlarged the size of the dataset. However, transcripts as video descriptions only work for very limited types of videos - like instruction videos, which makes the dataset not representative of real-world semantic diversity and complexity. To fix the problem of poor semantic correlation between the text and video, recent works [36, 42] have started to deploy automated video captioning and ASR together to make more detailed and comprehensive text descriptions. The result of such works proved that providing more correlated details on the text side does improve overall language-video representation and benefit downstream tasks.

However, videos are much more complex than just visual and speech. As the expectation on the language-video representation gets higher, more multi-modal information is extracted from the video representation side. On the other hand, the basic visual captioning and speech information on the text side failed to provide matching information on a more subtle or detailed level. This could then become the source of the next round of poor semantic correlation. **Can there be a solution to construct text descriptions that highly correlate to the vast variety of information from the video side in a flexible and automated manner?** Answering this question could be the key to achieving a more robust and holistic language-video representation. To tackle the issue, this work focuses on three main components:

Firstly, we introduce a new method that can be used to improve any existing video-language dataset to be more modality and context-aware. At the core of this method is a multifacet video captioning method to capture not only the basics such as the entities, actions, and speech transcripts but also the important and subtle aesthetics and emotional cues from video. The design of this automated captioning method ensures a flexible coverage of different aspects of the video and aims to always provide a matching amount of information compared to the video representation side.

Secondly, we compared different strategies for using the language model to create high-quality and factual textual descriptions and constructed an agent-like strategy to utilise the language model to ensure the text descriptions are of requirement and hallucination-free. Incorporating the language model also minimises human intervention in the process and ensures this can be applied to the dataset at scale.

Thirdly, we assess the feasibility and effect of such a method on overall language-video embedding through an important downstream task text-video retrieval. In the specific scope of this work, we use the early but open and representative MSR-VTT as the base dataset, and several representative multi-modal text-video retrieval models without pretraining to better understand the change brought by the dataset improvement.

Fig 4 provides an overview of our customised approach for the entire work. The lower half of the figure illustrates our methods that can be used to improve any existing video-language dataset to be more modality and context-aware. The upper half of the figure illustrates the overall evaluation plan using the downstream task of text-to-video retrieval.

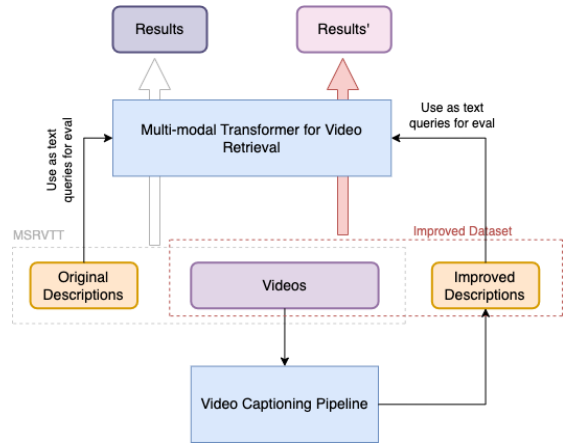

Fig. 1. A global view on our approach

## 2 RELATED WORK

### 2.1 Language-video Datasets

Language-video data pairs are always the key foundation to enable cross-modal representation and understanding. There are roughly three phases of language-video datasets development. In the initial phase, datasets are manually annotated and small. One example is MSR-VTT[41], an open domain dataset with 10k video and text pairs from human annotators for what roughly happens in the video. Another frequently used one is focused a bit more on the storytelling perspective, LSMDC[30] contains about 202 movies with either movie script or from transcribed DVS (descriptive video services). More domain-specific datasets like YouCook2[46] contain 2000 long untrimmed videos from 89 cooking recipes; on average, each distinct recipe has 22 videos.

Progressing with the maturing ASR technologies, as well as the popularised pretraining method, language-video datasets in the second phase are getting much larger, and have started to deploy automatic methods for annotation to cover the large-scale pretraining needs. HowTo100M[24] constructs such pairs using the video and their narration. Using ASR greatly reduces the cost, and it features a total of 136M video clips with captions sourced from 1.2M YouTube videos. This opens the door to large-scale language-video datasets and enables pretraining beyond text-image (such as CLIP). However, these large-scale datasets suffer from issues brought about by the low correlations in semantics between the language and the video. In many other cases aside from instructional or tutorial videos, ASR-captured speech or narration texts may not reflect what is happening visually on the screen.

In the most recent phase, works start to explore ways to incorporate more automatic tools and improve the semantic alignment of the text descriptions and the videos. Recent works [36, 43] have started to incorporate ASR, image captioning tools, and language models to automatically create more correlated text descriptions for the video. However, these new attempts only focus on the coarser level description of key events happening in the video visually and the speech aspect. More intricate details and subtle nuanced elements such as aesthetics and emotional cues are not touched at all.

## 2.2 Video Captioning

Video-to-text (V2T) technology has evolved significantly in recent years, driven by advancements in deep learning and artificial intelligence[6]. V2T models have achieved impressive results on a variety of benchmarks. They can be used for a wide range of applications, such as video captioning and video retrieval.

The field of video captioning has experienced growing interest from both computer vision and natural language processing researchers[28], as it aims to generate text descriptions of video content. Some previous works concatenate different visual features to obtain the video description [34, 45].

Another approach consists of using multi-modal pretraining models. The work of Ging et al. [12] involves different levels of granularity and modalities to build the Cooperative Hierarchical Transformer (COOT), which includes an attention-aware feature aggregation layer that utilizes the local temporal context. This method improves the performance of the video captioning model but does not consider some important modalities such as background sounds and conversations. VideoBert[32] and ActBert[47] employ on off-the-shelf automatic speech recognition (ASR) models to gather large repositories of paired video sequences and textual annotations. Both models utilize a BERT-style objective function to simultaneously train the video and text encoders.

To obtain more detailed descriptions, some works seek help from language models, Seo et al. proposed an end-to-end generative pretraining model [31] which is able to generate sentences based on the multi-modal context. They utilize the future utterance as an additional text source to address the absence of captions in unlabeled videos. Other works, such as [4, 5], focused on directly applying language model-based workflow to annotate videos in more comprehensive detail. While adding more details to the descriptions, these end-to-end methods tend to focus still solely on visual modality and sometimes encounter the inevitable hallucination problem.

## 2.3 Multi-modal Text-video Representation

Oftentimes, video-and-language representation utilizes contrastive learning to align video and text features within a joint latent space based on their semantic similarities. Aside from the spatio-temporal features, videos also encompass attributes in various modalities such as audio, motion, and Optical Character Recognition (OCR), and motion information.

Some works, such as MMT[10] and MDMMT[15], have studied the importance of including multi-modal video features. Different "Experts" are incorporated for these modalities, with each "Expert" being a specialized encoder to extract features from a certain modality. In these works, the extracted multi-model features on the video side are combined with the spatial-temporal features to enhance the semantic representations. More specifically, the different modality features can include scene embeddings, face features, motion features, audio features, and Optical Character Recognition (OCR).

However, with the more detailed features from the video side, video-and-language representation is facing a mismatch in fundamental aspects: the text features that are used to align with the video features are often lacking in semantic richness. Most of the methods and datasets are still pairing videos with plain and basic text descriptions.

Therefore, we would like to bridge this gap by introducing a robust way to build a more semantically aligned dataset with more holistic narrative elements and details. As explained in the work of Pérez-Mayos [25], by optimizing the annotation of the training material, the quality of video-and-language representation can be improved.

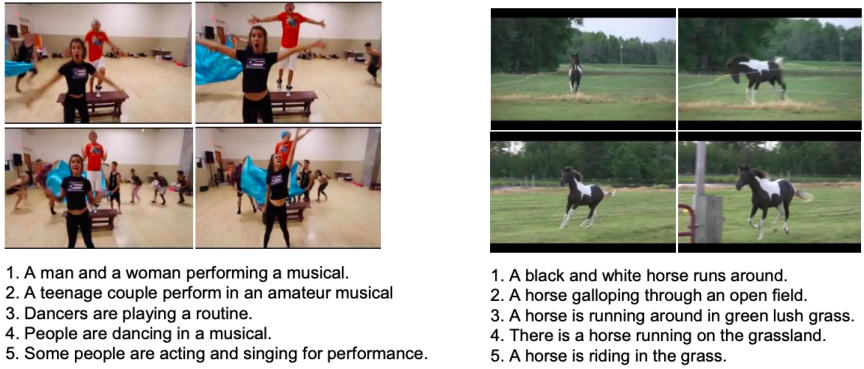

Fig. 2. Examples of the videos and annotations in MSR-VTT. Two samples are given, each containing four frames to represent the video clip and five human-labeled sentences

## 3 METHOD

### 3.1 The MSR-VTT Dataset

The base dataset we used for this work is MSR-VTT (Microsoft Research Video to Text). It is a classic language-video benchmark proposed by Xu et al. [41] in 2016. It was compiled by gathering over two hundred popular search queries from YouTube, with over one hundred videos associated with each query. MSR-VTT offers in total 10,000 video clips each lasting 10 to 20 seconds. Every clip is annotated with approximately 20 natural sentences by Amazon Mechanical Turk. It sums up to 200,000 sentences for all the videos, making it one of the most classic datasets in terms of both sentence and vocabulary count. Covering 20 diverse categories, the videos span a wide spectrum of topics to represent the real-world situation fairly (Fig ??). However, as can be seen in Fig ??, the annotation vastly focuses on the plain visual happenings in the video. We consider this obvious lack of holistic details in the description to be a great opportunity to test and evaluate our method.

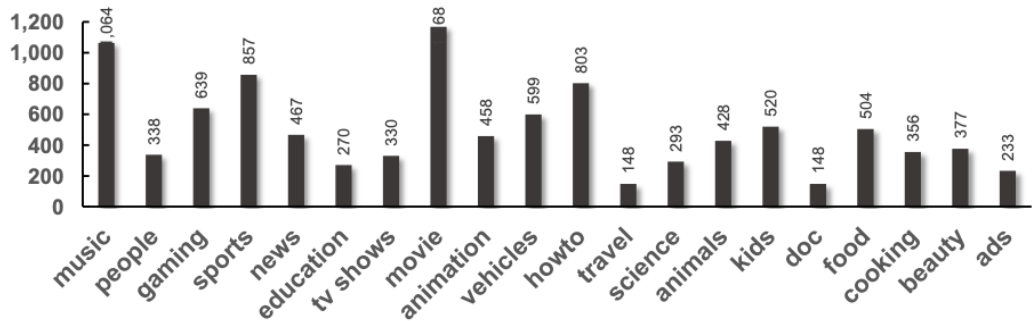

Fig. 3. The distribution of video categories in MSR-VTT dataset [8].

We obtained the dataset following the instructions from the work of Li et al. [17]. Then we followed the common protocol[16, 24, 44] and employed the same data split as [17], which comprises 7010 clips and associated texts for training and validation, and the results were reported on 1k video and description pairs sampled from the test set of MSR-VTT. It is worth noticing that upon inspection, we discovered that 834 videos in the training and validation set lacked audio, as well as for 354 videos in the testing set. We excluded these videos from our analysis to maintain data consistency and ensure the integrity of our experiments.

## 3.2 Multifacet Video Captioning

Whatever type of videos we work with, they are essentially an audiovisual story. The most important narrative elements in a story consist of plot, dialogue, setting, characters, point of view, (emotional) tone, and style [2]. They are crucial for the understanding and navigation of video content. Despite the significant progress in video captioning techniques, existing language-video dataset often fails to capture most of these elements. Recent works, such as InternVid [36], using language models to generate more in-depth and diverse annotations still fail to come about this task from the narrative perspective - either coming up with arbitrary new details or not controlling the model's hallucinations. To address these limitations and maintain the benefits of multi-modality, we propose a new video captioning pipeline that focuses on and is guided by the key narrative aspects, with the help of language models (see Table ?? for a comparison).

| Dataset | plot | setting | characters | point of view | dialogue | tone | style |
|---|---|---|---|---|---|---|---|
| MSR-VTT | ✓ | | ✓ | | | | |
| InternVid | ✓ | ✓ | ✓ | | limited | | |
| Holistic-MSR-VTT (Ours) | ✓ | ✓ | ✓ | | ✓ | ✓ | ✓ |

Table 1. Summary of fundamental narrative elements integrated in different datasets

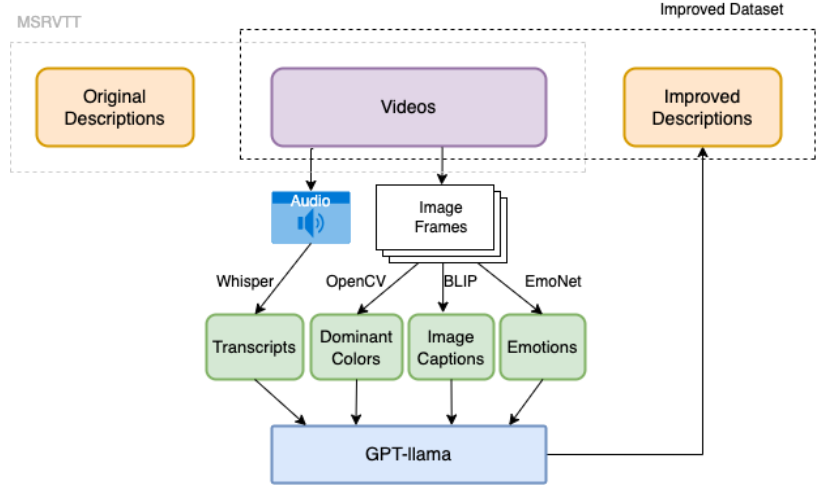

Fig. 4. Our video captioning pipeline

We specifically paid attention to intricate and nuanced details to echo the tone and style aspects. For this, the pipeline utilizes separate models to extract key elements that are essential for constructing the holistic description from different modalities and aspects of the video, including visual, dialogue, tone, and style captioning.

By employing this multifaceted approach, we aim to achieve a more nuanced and grounded video captioning system that effectively captures the essence of the video while maintaining control over the generated descriptions. In addition, the pipeline's modular architecture makes it adaptable to future modifications. We can effortlessly integrate new video aspects or remove those that are no longer relevant, enabling the pipeline to evolve in accordance with changing video captioning requirements.

We carefully selected the state-of-the-art models for each keyword extraction task in our pipeline, considering the availability of source code and the computational resources at our disposal. The detailed experts used for the key elements in the captioning are listed in the following sections.

*3.2.1 Visual Captioning.* The task of getting the main visual annotation covering the plot, setting, and characters is converted into a sequential image captioning task. There are many choices for image captioning, such as state-of-the-art close-set segmenters like Segformer[40] and Oneformer[39], which efficiently provide detailed semantic descriptions. CLIPSeg[23], an open-set segmenter, goes beyond conventional categories, generating more creative captions. Also SSA (Semantic Segment Everything) [3] distinguishes itself by segmenting and classifying every pixel, yielding exceptionally detailed descriptions.

Our first attempt was to use SSA since it inherits from SAM (SegmentAnything Model) [13] the powerful ability of arbitrary object segmentation, and also provides dense open-vocabulary category annotations. However, SSA seems to be unsuitable for captioning tasks with images that have intricate backgrounds or excessive detail, leading to captions that are dominated by background elements and fail to adequately capture the main objects in the scene 5.

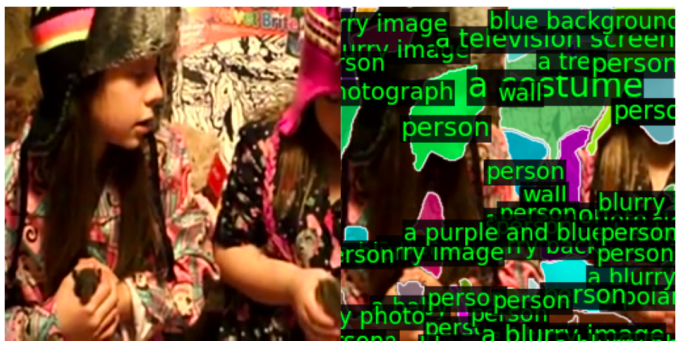

Fig. 5. Example of segmentation using SSA

Then we decided to utilize the BLIP model to describe our image sequence frames. BLIP [18] introduces a pretraining Vision-Language Processing (VLP) framework that harnesses the power of noisy web data to build robust, generalizable, and data-efficient models. By leveraging a dual-stream architecture and a three-stage pre-training process, BLIP effectively fuses visual and linguistic information, achieving superior performance on a variety of VL tasks compared to traditional methods. BLIP's ability to learn from diverse data, including noisy web sources, demonstrates its robustness to real-world variations, while its data efficiency streamlines the training process and reduces computational costs. These advantages position BLIP as a promising framework for advancing VL research and enabling machines to bridge the gap between vision and language with unprecedented accuracy and robustness.

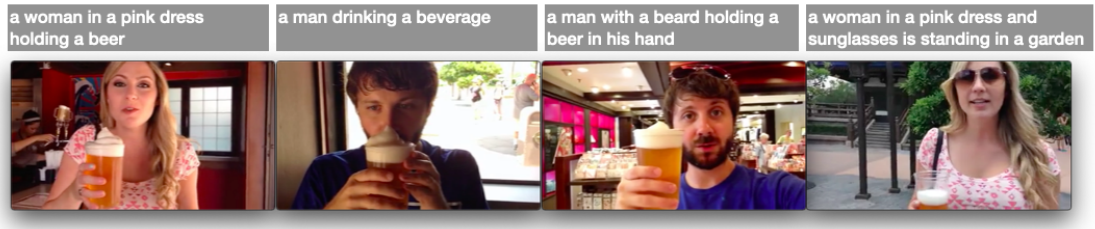

Fig. 6. An example of key sentences generated by BLIP for extracted frames of a video from MSR-VTT

We extract four frames per second from the video and employ the BLIP model to identify objects and provide a comprehensive overview of the visual content of each frame. This approach enables us to generate a sequential set of descriptions, each corresponding to a specific frame (see Fig 6 for examples).

*3.2.2 Dialogue Captioning.* Whisper[27]: propose a novel approach to speech recognition that utilizes large quantities of unlabeled speech data and weak supervision signals. This approach stands in contrast to traditional supervised learning methods that require carefully curated datasets of labelled audio and text pairs. By leveraging weak supervision, the proposed model is able to learn from a much broader range of data, including noisy and uncurated sources, leading to improved robustness and generalization capabilities.

The core of the proposed model is a transformer-based architecture that is trained to predict the most likely transcript for a given audio clip. The training process is guided by weak supervision signals, which consist of coarse alignment information between audio and text segments. These alignment signals provide the model with a general sense of the temporal correspondence between speech and text, allowing it to learn linguistic patterns and acoustic features without the need for explicit transcription labels.

The proposed approach offers several advantages over traditional supervised learning methods. First, it enables the training of robust speech recognition models from a vast amount of unlabeled data, which is significantly more accessible than labelled data. Second, weak supervision relaxes the need for manual data curation and annotation, making the training process more efficient and scalable. Third, the model's ability to learn from diverse and noisy data leads to improved generalization performance, making it more robust to real-world audio conditions.

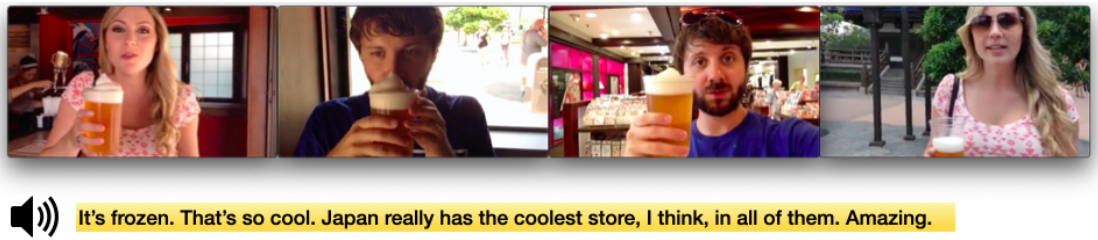

Fig. 7. Example of transcription of the dialogue in a video clip

To transcribe multilingual audio into English, we first extract the audio from the video using AudioSegment[1]. Then, we employ Whisper[27] to identify the language of the audio and transcribe it with the provided "small model". If the original language is not English, we set the task parameter to "translate" which enables the transcribe function to generate directly an English transcription of the foreign language audio (see Fig ?? for examples).

*3.2.3 Tone Captionning.* In the context of audiovisual understanding, the tone usually indicates the overall emotional cue within the content ??. In this work, we implement EmoNet?? for tone detection in videos. It is a deep learning-based framework for estimating continuous valence and arousal levels from faces in naturalistic conditions. this approach utilizes a 3D CNN to extract features from facial images and an RNN to model the temporal dynamics of facial expressions. We are able to recognize 8 key emotions, including neutral, disgust, happy, anger, contempt, sad, fear and surprise. Our approach for this model consists of extracting 2 frames per second and annotating the emotion of each frame, then counting the total number of times for each emotion to appear. Finally, the most frequently appeared emotion as used as the base tone for the video.

*3.2.4 Style Captioning.* The stylistic aspect of audiovisual content is complex. Since this work's focus is not to fully capture the stylistic features but to showcase the improvement of a more holistic language-video dataset would bring to the overall representation task. Also because of the modular design of our method, it would be easy to add more details. For the ease of the experiment, we choose the dominant colour as the stylistic feature to work with, as it is one of the keys for representing style in audiovisual language [11]. For this, we employ the K-means clustering algorithm, a powerful tool for grouping data points based on similarity. We leverage the OpenCV library to effectively extract and analyze colour information.

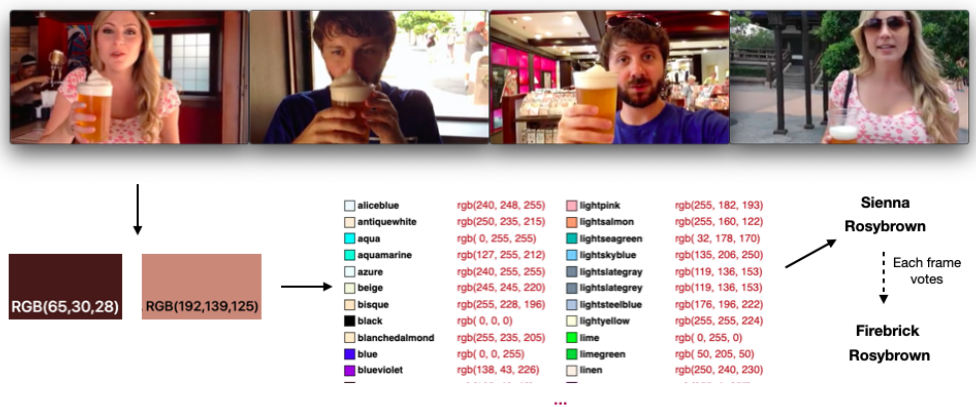

Fig. 8. Process of dominant colour detection based on one frame of the example video

In our case, after extracting two frames per second from the video, we use K-means to cluster the pixels of each image into two clusters, which will represent the dominant colours of the image frame in RGB format. Next, we utilize a colour name lookup table package [2] that maps the RGB value set into 147 colour names. It helps us match the dominant colour code to its nearest colour name. Finally, we compute the occurrences of dominant colours across all frames and identify the top two colours, which are considered the dominant colours of the video.

---
[1]https://github.com/jiaaro/pydub
[2]https://pypi.org/project/webcolors/

| Strategy | Pros | Cons |
| --- | --- | --- |
| Basic | Fast and straightforward. | Lack of language consistency and can be arbitrary results. |
| Role-play | Allows for more detailed and nuanced prompts. Can be effective for generating dialogue or narratives. | Can be overly creative to invent nonexistent details. |
| Template | Provides structure and guidance for the model's response. Ensures consistency and reduces randomness in outputs. | May lead to formulaic and repetitive outputs if not designed well. |
| Rule | Offer fine-grained control over the content and style of the output. Can be used to avoid unwanted elements or biases. | May limit the model's creativity and expressiveness. |

Table 2. Summary of Prompt strategies

The significance of using K-means clustering for colour extraction in video captioning lies in its ability to provide insights into the visual theme of the video. By identifying the dominant colours, we can capture the essence of the video's aesthetic and incorporate this information into the generated captions.

### 3.3 Language Model Enhancement

Using language models to directly generate annotations for audiovisual content can be extremely costly and sometimes hard to align with factual information. Instead, the language model in this work is used as an agent. With the visual, dialogue, tone, and style annotations obtained, we resort to using language to automatically combine them to generate a comprehensive and accurate video description.

Llama 2 [7] is a trained large language model boasting impressive capabilities in conversation and text generation. Developed by Meta AI, it aims to bridge the gap between large, closed-source models and more readily available yet less powerful ones. Unlike many language models, Llama 2's code and pre-trained weights are entirely open-source, allowing researchers and developers to freely access and modify it. This fosters transparency and collaboration within the AI community. Furthermore, Llama 2's multilingual capabilities facilitate cross-cultural understanding, emotional intelligence fosters empathy, and domain adaptability caters to specific tasks. Built on the foundation of its predecessor, Llama 1, Llama 2 surpasses it in many aspects. It demonstrates strong performance in conversational tasks like question answering, text generation, and summarization. In our study, we employ the fine-tuned 7B chat model of Llama 2 [3] to combine the multifaceted annotations extracted by expert models in the pipeline and generate the improved caption for MSR-VTT.

Prompting plays a crucial role in shaping the behaviour of generative language models and can significantly impact the quality, reliability, relevance, and even bias of the generated text. Therefore, it is essential to carefully consider different prompt strategies when employing language models for video captioning tasks. Various prompt strategies can be employed to enhance video captioning. A well-chosen prompt can give the language model a richer context, leading to more comprehensive and informative captions [37]. Table 2 lists the other strategies we tested.

---
[3]https://github.com/meta-llama/llama/blob/main/MODEL_CARD.md

For each of the strategies, we give the same exemplary annotation results obtained from the previous pipeline:

- *Information Chunk*: "'[visual: "(frame01: a woman in a pink dress holding a beer, ..., frame15:a man drinking a beverage,...)]"] [dialogue: "It's frozen. That's so cool. Japan really has the coolest store, I think, in all of them. Amazing." ] [tone: "surprise"] [style: "firebrick, rosybrown"]"'

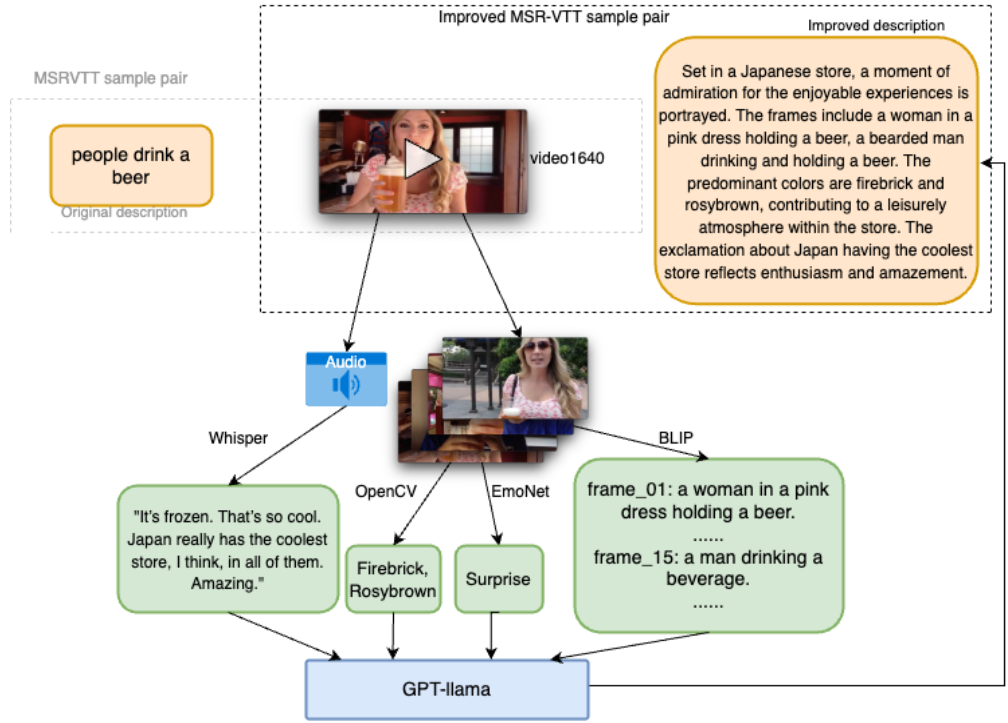

Fig. 9. Our video captioning pipeline illustrated through an example from MSR-VTT

*3.3.1 Basic Example.* We deployed a basic strategy to serve as a benchmark to be compared with other trendy strategies that are more sophisticated. This basic strategy is a zero-shot learning approach [14] that simply gives the model the raw results of the mentioned models for each video as input without any example or detailed instruction. When feeding the information chunk and asking the model to "write me a video description based on the information chunk", the language model produces the following result:

- *Output*: The video appears to be showing a person holding a beer and smiling, with the background showing a frozen scene. The person in the video seems to be expressing surprise and excitement, with a tone of amazement. The style of the video is reminiscent of a warm and inviting firebrick color, with a touch of rosybrown. Based on the dialogue provided, it seems that the person in the video is commenting on the coolness of a store in Japan. The use of the word "coolest" and the exclamation "Amazing!" suggests that the person is very impressed with the store. Overall, the video seems to be showcasing a fun and exciting shopping experience in Japan, with a focus on the unique and cool products available at the store.

*3.3.2 Role-play Example.* On top of just feeding the information to the model, following [1], we dived an experiment on the role-playing strategy for the writing task in the hope of a more detailed and professional description. we crafted a persona and instructions that could be suitable for the task:

- *The Narrator*: Assuming the role of an impartial narrator at a cultural heritage institution. your job is to provide detailed and objective descriptions of certain videos to the audience in language because they cannot access them. Your goal is to make the description easy to understand, logical, holistic in detail, and not too lengthy using the information provided in the information chunk for each video.

An example of the result from such a prompt is illustrated here:

- *Output*: The video unfolds in a Japanese store, hailed as the coolest, as the narrator expresses genuine surprise at the frozen spectacle. The tone reflects an overarching sense of amazement. The image sequences offer glimpses into the vibrant scene – a woman donning a striking pink dress and sunglasses captivates the frame, holding a beer, contributing to the overall laid-back atmosphere. Another frame captures a man sipping a beverage, while a bearded individual holds a beer, adding diversity to the dynamic setting. The chosen color palette enhances the visual experience, with firebrick and rosybrown dominating the scene, creating a warm and cozy ambiance. The narrator's enthusiasm is palpable, evident in the exclamation about Japan having the coolest store. The combination of expressive transcriptions, emotional tone, and color detectors paints a vivid picture of a unique and enjoyable experience within the confines of this exceptional Japanese establishment.

*3.3.3 Template Example.* While few-shot learning with language models shows promising results for better results, the cost in terms of tokens and time can be significant[20]. To address these limitations, we explore another prompting strategy that leverages pre-defined templates[29]. By providing a language model with a structured framework, we aim to achieve comparable accuracy while reducing computational demands and streamlining the process.

Here is an example using a template with the same video information:

- *Prompt*: Please generate a video caption following this template: "'The video painted in hues of ["style 1"] and ["style 2"], begins with [first few frames in visual annotation]. Then, [middle frames in visual annotation] can be seen. We hear [dialogue] which means this video is talking about [Interpretation of the dialogue]. The video concludes with [last few frames in visual annotation]. We are left with a feeling of [tone annotation].

With the same information chunk provided in the role-play example, the result is as follows:

- *Output*: The video, painted in hues of firebrick and rosybrown, unfolds with the scene of a woman in a pink dress holds a beer, setting the stage for an surprising journey. The narrative is filled with awe as a man indulges in drinking a beer, and holds the beer in his hand. The visual aesthetics are heightened as a woman in a pink dress and sunglasses stands in a garden.The transcript adds another layer of fascination: 'It's frozen. That's so cool. Japan really has the coolest store, I think, in all of them. Amazing.' The surprises continue to unfold, creating a sense of wonder and amazement. The video concludes with a feeling of surprise, leaving us captivated by Japan's unique charm shines captured within the vibrant frames.

*3.3.4 Rule Example.* Rules are important for writing tasks for language models. Following the suggestions from [19], we resort to explicit rules to apply to the prompt when doing this task.The aim is to hinder the problem of hallucinations and prevent the model from creating non-factual details. The prompt given is as follows:

- *Prompt*: Please generate a video caption adopting a third-person point of view. You can only incorporate the visual, dialogue, tone, and style annotations provided in the following information chunk and not other improvised details. In the chunk, Visual annotation is the frame-by-frame descriptions of visual happenings in the video; Dialogue annotation is the speech happening in the video; Tone annotation is the general emotional tone of the people; Style annotations is the overall colour of the video.
- *Output*: Set in a Japanese store, a moment of admiration for the enjoyable experiences is portrayed. The frames include a woman in a pink dress holding a beer, a bearded man drinking and holding a beer. The predominant colors are firebrick and rosybrown, contributing to a leisurely atmosphere within the store. The exclamation about Japan having the coolest store reflects enthusiasm and amazement.

After comparing the outcomes of each prompting strategy, we determined that employing the rule strategy would deliver the most desirable results for our video caption generation task as it is in the right length and with the proper amount of factual details. An overview of the whole improved annotation pipeline is illustrated in Fig 9 with examples.

## 3.4 Qualitative Results for Improved Annotations

In this section, we illustrate the improvement of the annotation with some qualitative results. In the following examples of captioning, we observe that the improved descriptions are indeed richer and contain more information that could be potentially interesting for the users.

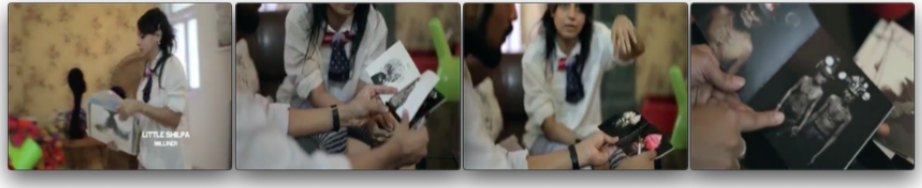

Fig. 10. Video clip in MSR-VTT with two original and two improved descriptions

For example in Fig 10, upon inspection, the content of the conversation was successfully integrated into the caption of the video and we gained extra knowledge about the photo album displayed in the video. We could not have known that it was about artists in Bangalore with the original MSRVTT annotations.

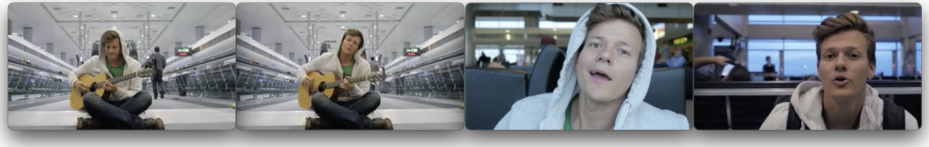

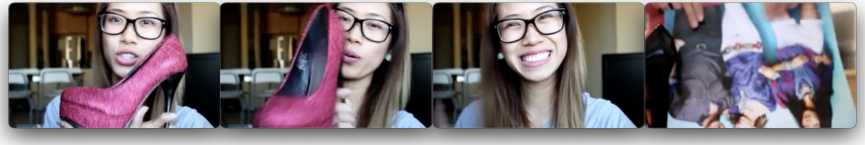

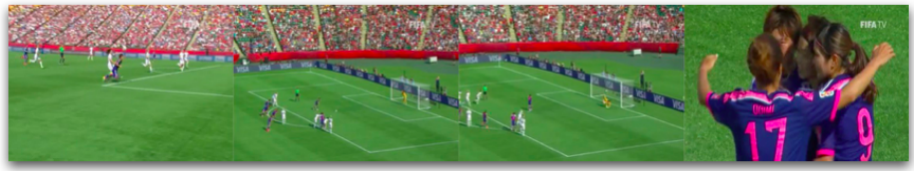

Fig. 11. Video clips in MSR-VTT with one original and one improved description, details that are added are shown in highlight.

Furthermore, in the case of Fig 13 our pipeline effectively captured the colours and emotions depicted in the video. The colours representing the football teams aided in understanding the plot, and the characters' joyful expressions were meticulously captured. These fundamental narrative elements, introduced by the enhanced dataset, constructed a cohesive storytelling of the video, absent in the manually annotated sentences of the original dataset.

## 4 EXPERIMENT

We select the popular downstream task, text-to-video retrieval, to test the actual improvement of a more detailed and holistic annotation in the training dataset for language-video representation learning. The experiments are conducted based on the original and improved MSR-VTT dataset. More details for the experiment setup are listed in the following sections:

### 4.1 Model Selection

This work assumes that with more information becoming available through multi-modal expert encoders on the video side, the whole language-video representation learning process is hindered by the lack of correlating information on the text side. The assumption is that regardless of the detailed strategies used for the training process, the more detailed and matching information from the text side would improve the overall training.

For this, we have ignored models that do not stress the multi-modal features from the video side and only focused on the multi-modal models. Most multi-modal models focus on extracting multi-modal representations with different extractors (also known as experts) and then fuse them to obtain the final video features for matching with the text side features. On the other hand, we want to evaluate and compare the results brought by only the improvement presented in the new language-video pairs, so we purposefully ignored the models that depend on large-scale pretraining to emphasise that. On top of these, based on the code availability, we chose MMT[10], HCQ[33], and T2VLAD[35] for our experiment. The three models all have multiple experts on the video representation side and have different alignment and training strategies, which make them perfect for testing our hypnosis.

|  | Original pairs query | Improved pairs query |
|---|---|---|
| Model trained with original pairs | [1] | [2] |
| Model trained with improved pairs | [3] | [4] |

Table 3. Benchmark set up for each retrieval model

### 4.2 Benchmark Set Up

We adopt the "1K-A" split for the training and testing originally proposed in [44]. We kept this split in both the original and the improved MSR-VTT for our experiment. For each of the selected models, two versions are trained. One trained on the original MSR-VTT language-video pairs, and one trained on the improved MSR-VTT pairs. We use the same parameters as proposed in respective papers and their official implementation GitHub repository.

To remain consistent between all models, we use the three widely used metrics r@1, r@5 and r@10 to access the performance of the retrieval models. A detailed plan for an experiment with one model is showcased in Table 3. As pointed out in [38], such metrics are indicative based on the assumption that only a single caption is relevant to a query video and vice versa. As descriptions made for videos are simple and general, this assumed one-to-one relationship might be problematic. However, as the language-video pairs become more detailed and explicit, the indicator might be more representative. To understand the performance of the language-video representation learned using the original and the improved pairs, we calculate the metrics using the original test pairs and the improved pairs. This is also to understand which representation space would handle the more complex and diverse query reality better.

By comparing [1] with [3] and [2] with [4], we can conclude if the additional information is useful for improving the model's performance in simple, or complex query situations. Furthermore, comparing [1] and [4] will provide some additional help in understanding the change in performance in situations where queries are matching to the training situation. Comparing the changes across different models can help us to conclude the robustness of such a method in language-video representation and not model-specific, i.e. specific to one training strategy for representation learning.

## 4.3 Comparison of the performances

|                      | MSR-VTT query |      |      | Improved query |      |      |
|----------------------|---------------|------|------|----------------|------|------|
| Model                | r@1           | r@5  | r@10 | r@1            | r@5  | r@10 |
| MMT × MSR-VTT        | 24.5          | 54.0 | 67.7 | 10.0           | 26.3 | 39.7 |
| MMT × Improved       | 24.1          | 53.2 | 63.5 | 35.2           | 60.4 | 72.0 |

Table 4. Performance of MMT retrieval models for r@x

|                      | MSR-VTT query |      |      | Improved query |      |      |
|----------------------|---------------|------|------|----------------|------|------|
| Model                | r@1           | r@5  | r@10 | r@1            | r@5  | r@10 |
| HCQ × MSR-VTT        | 25.9          | 54.8 | 69.0 | 13.1           | 27.4 | 40.1 |
| HCQ × Improved       | 26.1          | 53.7 | 68.9 | 36.9           | 60.8 | 77.2 |

Table 5. Performance of HCQ retrieval models for r@x

|                      | MSR-VTT query |      |      | Improved query |      |      |
|----------------------|---------------|------|------|----------------|------|------|
| Model                | r@1           | r@5  | r@10 | r@1            | r@5  | r@10 |
| T2VLAD × MSR-VTT     | 29.5          | 59.0 | 70.0 | 13.4           | 29.5 | 40.3 |
| T2VLAD × Improved    | 28.8          | 59.2 | 70.4 | 37.4           | 61.3 | 77.6 |

Table 6. Performance of T2VLAD retrieval models for r@x

The result from the proposed experiment is presented in Table 4, Table 5, and Table 6. When evaluated with the original queries from the MSR-VTT test set (compare [1] with [3]), we can see the model trained on the improved pairs performs very much comparable with the model trained on the original pairs. Although ideally, we would prefer a more obvious increase in this category. However, as mentioned in the previous section, since the assumed one-to-one relationship between the simple and general query and the video descriptions does not always reflect the real performance of the retrieval tasks on more subtle levels, these comparable results are expected.

This also shows that providing more information from the text side correlated to the video side would not hinder the performance, such information is not noise. This is further proved by the overall improvement indicated in the comparison in a more complex query setting, where queries are from the improved pairs (compare [2] with [4]). As the query becomes more complicated and detailed, models trained on the original pairs fail to perform at the same level. However, when the model is trained with improved pairs, an evidential boost in performance can be observed.

It can be argued that using the text queries native to its training set (for example, the model trained on the original dataset will perform better for original queries as opposed to the model trained on the improved dataset, since the query is similar to the training data) is part of the reason for the performance difference. However, the comparison between [1] and [4] shows that even in the native query setting, the improved dataset still provides a better embedding space for retrieval tasks.

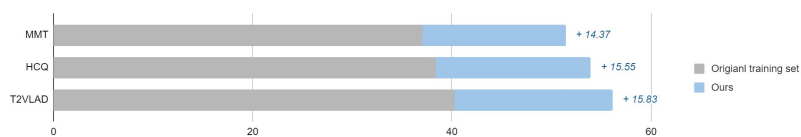

Fig. 12. Performance differences brought by training using the improved MSR-VTT dataset across different models. The difference is highlighted in blue.

We calculate the average performance difference between the two versions of the model in all query settings and all r@x across models. This indicator can help us to understand if the more detailed and correlated text descriptions can only improve the language-video representation with a specific training strategy. As indicated in Fig 12, our proposed method is applicable and effective to all the models we selected. This means having the text side information as detailed and correlated to the video side information is helpful and necessary before considering the model or optimisation architecture for training a language-video representation space. In Fig 13 we introduced qualitative examples for the different retrieval results for references.

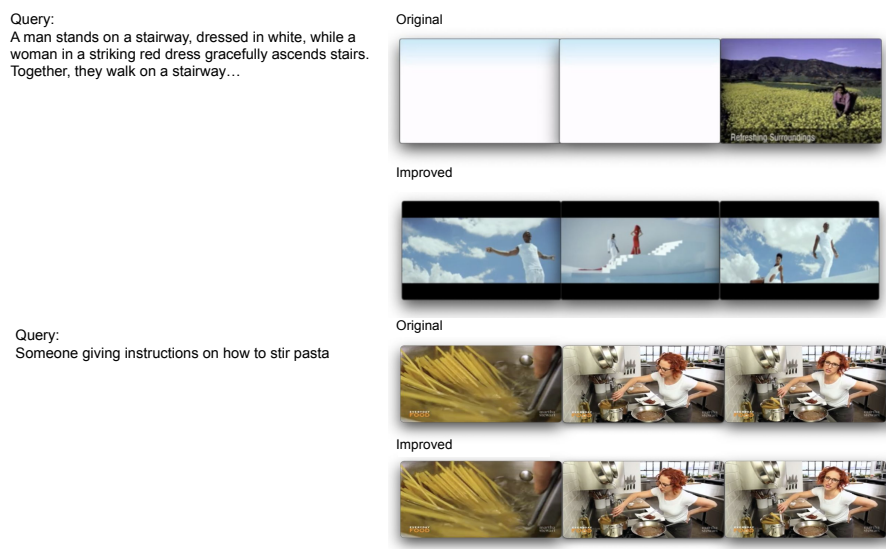

Fig. 13. Retrieval examples with text query: In each example, the upper plot showcases the first result returned by the MMT model trained with the original MSR-VTT dataset, and the lower plot corresponds to the result returned by the MMT model trained with improved MSR-VTT dataset. The lower example shows the result of retrievals with a simple query (original MSR-VTT text), while in the upper example, the improved MSR-VTT text is used as a query.

## 5 CONCLUSION AND FUTURE WORK

In conclusion, this paper presents advancements in the realm of video-language datasets and their corresponding applications. Firstly, we introduced a novel method to enhance existing video-language datasets, making them more modality and context-aware. This multifaceted video captioning method captures a wide range of elements, including entities, actions, speech transcripts, aesthetics, and emotional cues, ensuring a comprehensive and balanced representation of the video content. This modular design also ensures the flexibility to make semantic details in the text descriptions correlate with the video representation and alignment used in various multi-modal models for training the language-video representation.

Secondly, we explored various strategies for employing language models to generate high-quality, factual textual descriptions. By developing an agent-like strategy that leverages the language model, we ensured that the descriptions meet the required standards and are free from hallucinations. This approach also reduces human intervention, enabling the method to be applied at scale.

Thirdly, we evaluated the feasibility and impact of our method on overall language-video embedding through the downstream task of text-video retrieval. Using the MSR-VTT dataset and several representative multi-modal text-video retrieval models without pretraining, we demonstrated the improvements brought by our enhanced dataset are significant and cross-model.

Despite these contributions, our study has certain limitations. The scope of our evaluation was restricted to the MSR-VTT dataset and a limited number of models. Future work should expand the size of the study to include a broader range of models and datasets, providing a more comprehensive understanding of the method's effectiveness. Additionally, our evaluation relied heavily on quantitative metrics such as recall at k (r@k), which may not fully capture the nuanced improvements in the dataset. Incorporating human evaluation in future studies will help address this potential bias and provide a more holistic assessment of the dataset's quality.

Overall, our work lays a robust foundation for further research and development in video-language datasets, offering promising directions for enhancing the richness and applicability of these datasets in various multimedia and AI applications.